\documentclass[twocolumn,english,aps,nofootinbib]{revtex4}
\usepackage[T1]{fontenc}
\usepackage[latin9]{inputenc}
\usepackage{amsmath}
\usepackage{amssymb}
\usepackage{appendix}
\usepackage{esint}
\usepackage{graphicx}
\usepackage{relsize}

\makeatletter

\newcommand{\p}{\boldsymbol{p}}
\newcommand{\xxi}{\boldsymbol{\xi}}
\newcommand{\F}{\boldsymbol{F}}
\@ifundefined{textcolor}{}
{%
 \definecolor{BLACK}{gray}{0}
 \definecolor{WHITE}{gray}{1}
 \definecolor{RED}{rgb}{1,0,0}
 \definecolor{GREEN}{rgb}{0,1,0}
 \definecolor{BLUE}{rgb}{0,0,1}
 \definecolor{CYAN}{cmyk}{1,0,0,0}
 \definecolor{MAGENTA}{cmyk}{0,1,0,0}
 \definecolor{YELLOW}{cmyk}{0,0,1,0}
 }

\usepackage{nicefrac}\usepackage{dsfont}
\usepackage[normalem]{ulem}\@ifundefined{definecolor}
 {\usepackage{color}}{}

\makeatother

\providecommand{\keyword}[1]{\textbf{\textit{Keywords: }} #1}

\usepackage{babel}
\begin{document}


\title{Mean first passage time of active Brownian particle in one dimension}
\author{A. Scacchi$^1$}\email {alberto.scacchi@unifr.ch}
\author{A. Sharma$^{1,2}$}
\affiliation{1. Department of Physics, University of Fribourg, Fribourg, Switzerland\\2. Leibniz-Institut f\"ur Polymerforschung Dresden, 01069 Dresden, Germany
}


\begin{abstract}
We investigate the mean first passage time of an active Brownian particle in one dimension using numerical simulations. The activity in one dimension is modelled as a two state model; the particle moves with a constant propulsion strength but its orientation switches from one state to other as in a random telegraphic process. We study the influence of a finite resetting rate $r$ on the mean first passage time to a fixed target of a single free Active Brownian Particle and map this result using an effective diffusion process. As in the case of a passive Brownian particle, we can find an optimal resetting rate $r^*$ for an active Brownian particle for which the target is found with the minimum average time. In the case of the presence of an external potential, we find good agreement between the theory and numerical simulations using an effective potential approach.
\end{abstract}

\maketitle

\keyword{active Brownian particles, mean first passage time, optimal resetting rate, effective equilibrium, escape process}


\section{Introduction}
Mean first passage time (MFPT) of a diffusing Brownian particle is a heavily researched problem. Recently MFPT has been studied in context of Active Brownian Particles (ABPs) which under go self-propulsion in addition to the Brownian motion~\cite{sharma2017escape}. An ABP performs persistent motion along the direction of an embedded unit vector which performs rotational diffusion. Often, the motion of ABPs is modelled on a coarse grained level by averaging out the orientational degree of freedom~\cite{farage2015effective,wittmann2017effective}, resulting in a non Markovian equation of motion. Following this approach, the authors in Ref.~\cite{sharma2017escape} studied the MFPT of an active Brownian particle diffusing in an external potential. In this paper, we preserve the orientational degree of freedom of the active particle and obtain its MFPT in 1-dimension. This allows us to verify the validity and accuracy of the coarse graining approach. We obtain the MFPT of an ABP in two scenarios, (a) freely diffusing ABP with stochastic resetting and (b) escape of an ABP over a potential barrier. For both these scenarios, we benchmark our findings with existing literature. 

Diffusion with stochastic resetting appears naturally in several search processes. A typical scenario would be to find a lost object with a random search strategy which is intermittently reset to its starting point. In Ref.~\cite{evans2011diffusion}, the concept of stochastic resetting was applied to a freely diffusing Brownian particle. It was shown that with a fixed target located at an arbitrary location, the resetting of the diffusion particle at a given rate, leads to a finite mean first passage time (MFPT). This is in contrast to a freely diffusion particle for which the MFPT is infinite. The most interesting finding of their work was that there exists an optimal resetting rate that yields the minimum MFPT. The concept of stochastic resetting has been extended to diverse fields such as search strategies~\cite{montanari2002optimizing}, microbiology~\cite{cates2012diffusive} and even the search and hopping behavior of capuchin monkeys~\cite{bartumeus2009optimal}. 

We focus our study on active Brownian particles (ABPs), a more general situation where Brownian particles are a special case. This generalisation allows a more flavoured and realistic description in various fields, such as those described above. For example, in a biological context, the constituting components of a system are active agents such as molecular motors inside animal cells. These motors are active particles which, in addition to the directed active motion, undergo Brownian motion\cite{molecular_motors}. Another example could be, on a macroscopic scale, the persistence of path in the context of the search of an object. While searching for a lost object, one persists on a certain time scale before resetting to the origin. Motivated by these considerations, we introduce activity as a key component of the motion of particle and study its impact on the MFPT.

We first investigate the MFPT of an ABP in 1-dimension with stochastic resetting using numerical simulations. The self propulsion in 1-dimension can be modelled by a two state model in which the particle moves with a constant velocity $v_0$ but switches the direction at a rate $\tau^{-1}$ in a random uncorrelated fashion. Introducing stochastic resetting to an ABP, as for a passive particle, leads to a finite MFPT. Moreover, one expects to find a corresponding optimal resetting rate for which the MFPT is minimum. The main question we focus on in this study is whether, one can reduce this problem of ABP to a passive Brownian particle with effective transport coefficients. For a freely diffusing ABP, this would correspond to an effective diffusion constant which is determined by $v_0$ and $\tau$. Can one express the optimal resetting rate in terms of this effective diffusion constant?

The mapping of ABPs to an effective equilibrium has been successful in describing the escape of an ABP over a potential barrier~\cite{sharma2017escape}. In the coarse grained approach, the stochastic equation of motion of an ABP includes a correlated colored noise and a white Gaussian noise. The strength and autocorrelation of the colored noise is determined by the activity parameters, the propulsion strength $v_0$ and $\tau$, the time scale of persistent motion of an ABP. We explicitly include the propulsion of ABP as a two state velocity and show that the MFPT of an ABP diffusing in an external potential can again be obtained as for a passive particle with effective diffusion constant and an effective potential; both of which are determined by the activity parameters.

\section{Model and theory}\label{model}
We consider a one dimensional system of a single Brownian particle of size $d$
with coordinate $x$ and orientation specified by an embedded unit vector $\p$. The orientation vector can point either along the positive $x$-axis or the negative one. The switching between these two states occurs at an average rate of $\tau^{-1}$ in an independent, uncorrelated fashion. A constant self-propulsion of speed $v_0$ 
acts in the direction of orientation. 
The motion can be modelled by the Langevin equation
\begin{align}\label{full_langevin}
\dot{x} = v_0\,\p  + \gamma^{-1}\F + \xxi
\end{align}
where $\gamma$ is the friction coefficient and the force on particle is generated from the external potential $\Phi(x)$ according to $\F\!=\!-\nabla \Phi$. 
The stochastic term $\xxi(t)$ is Gaussian distributed with zero mean and 
has time correlation	
$\langle\xxi(t)\xxi(t')\rangle=2D_t\delta(t-t')$ where $D_t$ is the translational diffusion coefficient. 

The stochastic resetting is implemented as a Poissonian process with a mean rate $r$ such that at every resetting event, the particle is put back to its starting point ($x = 0$).
When there is no external potential and no resetting, Eq.~\eqref{full_langevin} describes the well known dichotomous diffusion~\cite{balakrishnan2005connection}. The probability distribution of the particle can be obtained analytically in terms of modified Bessel functions. However, since we are interested in the MFPT, we consider the long time limit $t \gg \tau$, where it can be shown that the probability distribution reduces to a Gaussian with the diffusion constant 
\begin{equation}\label{effective_diffusion_coefficient}
D =  D_t+D_a=D_t+\frac{v_0^2 \tau}{2}.
\end{equation}

In order to perform analytics for a finite resetting rate $r$, we make the assumption that the motion of an active particle can be described at all time by the ordinary diffusion equation with the diffusion constant given in Eq.~\eqref{effective_diffusion_coefficient}. This is a reasonable assumption only under the self consistency condition that the MFPT is much larger than $\tau$. As shown below, this assumption holds when the target is placed sufficiently far from the initial position of the ABP, such that the MFPT is much larger than $\tau$. Under this assumption, the problem of finding the MFPT as well as the optimal resetting rate $r^{\ast}$ reduces to that presented by Evans and Majumdar in Ref.~\cite{evans2011diffusion}. Using the expression derived in Ref.~\cite{evans2011diffusion}, it follows that the MFPT $T(x_0)$ is given as 
\begin{align}
T(x_0) = \frac{1}{r}\left(\exp\left(x_0\sqrt \frac{r}{D}\right) - 1\right),
\label{mfpt}
\end{align}
 where $x_0$ is the location of the target object for an ABP which starts at the origin. 

The optimal resetting rate $r^{\ast}$ for an ABP freely diffusing in one dimension is given as
\begin{align}\label{optimal_rate}
r^{\ast} &=\frac{z^{\ast2}D}{x_0^2},
\end{align}
where $z^{\ast}$ is the solution of the transcendental equation $z^{\ast}/2=1-e^{-z^{\ast}}$.

This optimal rate minimizes the MFPT of a particle to a target object in a one dimensional system. The first aim of this work is to test Eq.~\eqref{optimal_rate} using Brownian dynamics simulations. In the second part of this work, we consider an ABP diffusing in an external potential. We particularly focus on the case of an ABP escaping a potential barrier. The escape problem is studied for the resetting rate set to 0. As in Ref.~\cite{sharma2017escape}, we choose the potential to have the form
\begin{equation}\label{potential}
\beta\Phi(x)=\frac{1}{2}\omega_0 x^2 - \alpha \mid x \mid^3,
\end{equation}
where $\beta = 1/k_{\rm B}T$, and $\omega_0$ and $\alpha$ are parameters. We are interested in the MFPT of a particle starting at the origin escaping over the barrier to be captured by a sink located sufficiently far from the barrier. In the effective equilibrium approach~\cite{farage2015effective,sharma2017escape} to ABP in an external potential, one obtains an approximate Fokker-Planck equation with an effective external potential $\Phi^{\rm eff}(x)$ and an effective position-dependent diffusion constant $D(x)$:
\begin{align}
D(x) &= D_t+\frac{D_a}{1+\frac{\tau D_t}{d^2} \beta \Phi^{\prime\prime}(x)}\, \label{Dx},\\
\beta \Phi^\text{eff}(x)&=\int_0^x dy\,\frac{\beta \Phi'(y)+D'(y)/D_t}{D(y)/D_t} \label{Vx}.
\end{align}
 
The effective diffusion constant is determined jointly by the activity and the potential and it reduces to the expression in Eq.~\eqref{effective_diffusion_coefficient} when there is no external potential. 

The expressions in Eqs.~\eqref{Dx} and~\eqref{Vx} were obtained under the assumption~\cite{farage2015effective,sharma2017escape,wittmann2017effective} that the stochastic process corresponding to time evolution of the orientation vector can be considered as a Gaussian noise process with a finite correlation time. In other words, one can disregard the explicit time evolution of the orientation vector of an ABP by including a colored noise term (exponentially correlated Gaussian noise) in the equation for the time evolution of the position of the ABP. There are two important conditions underlying this assumption: (1) The variance of the stochastic process corresponding to the orientation of an ABP decays exponentially in time independent of the spatial dimensions allowing the mapping to a colored Gaussian noise. However, the mapping is only approximate because the process is not Gaussian. This is most evident in one dimension where the process is a random telegraphic process and (2) the correlation time $\tau$ is sufficiently small such that the particle's displacement during this time period is negligible. This is required to ensure that there is no significant change in the potential that the particle experiences over a time period $\tau$.

 In Ref.~\cite{sharma2017escape}, the authors validated the theoretical predictions of the effective equilibrium approach against the numerical simulations of a stochastic process driven by colored Gaussian noise. The strength and autocorrelation of the colored noise were determined by the activity parameters $v_0$ and $\tau$. In our numerical simulations, we explicitly consider the orientation of the particle. Since we consider one dimensional system, the orientation can only take two values which is why we model it as a two state random telegraphic process. We will calculate MFPT and compare with the predictions of the effective equilibrium approach. For the sake of completeness, the expression for the inverse MFPT obtained in the effective equilibrium approach is given as~\cite{sharma2017escape}
 \begin{align}\label{MFPT_EFF_EQ}
 T^{-1} =r_\text{pass}\exp{\left({\frac{D_a\left(\beta E_0 - \omega_0\tau \right)}{D_t+D_a}}\right)},
 \end{align}
 where $r_{\rm pass} = \beta D_t\omega_0\exp(-\beta E_0)/(2\pi)$ is the escape rate of a passive particle over the potential barrier (Eq.~\eqref{potential}) and $\beta E_0 = \omega_0^3/(54\alpha^2)$ is the height of the barrier.

Before presenting the results, we give a brief description of the simulations performed in this study. We set the particle size to $d=1$. Time is always measured in units of $d^2/D_t$ and we have chosen $D_t = 1$ which together with $k_{\rm B}T = 1$ gives the friction coefficient $\gamma = 1$. The Eq.~\eqref{full_langevin} is integrated in time to generate the particle trajectory by advancing time in steps of $dt = 10^{-4}$ for the free diffusing particles and $dt=10^{-3}$ in the case with an external potential. The integral of the stochastic process $\xxi(t)$ over a time interval $dt$ is taken as a Gaussian distribution with zero mean and variance $2dt$. At every time step, the orientation of the particle is flipped with a probability $dt/\tau$. Similarly, stochastic resetting is implemented by resetting the particle to the origin with a probability of $rdt$ during every time step. The first passage time is calculated by averaging over  $10^5$ trajectories for a freely diffusing ABP. In the case of an external potential, the MFPT is obtained by averaging over $10^3-10^4$ trajectories.

\section{Results and discussion}

\begin{figure}[t]
\begin{tabular}{@{}c@{}}
\includegraphics[width=0.53\textwidth]{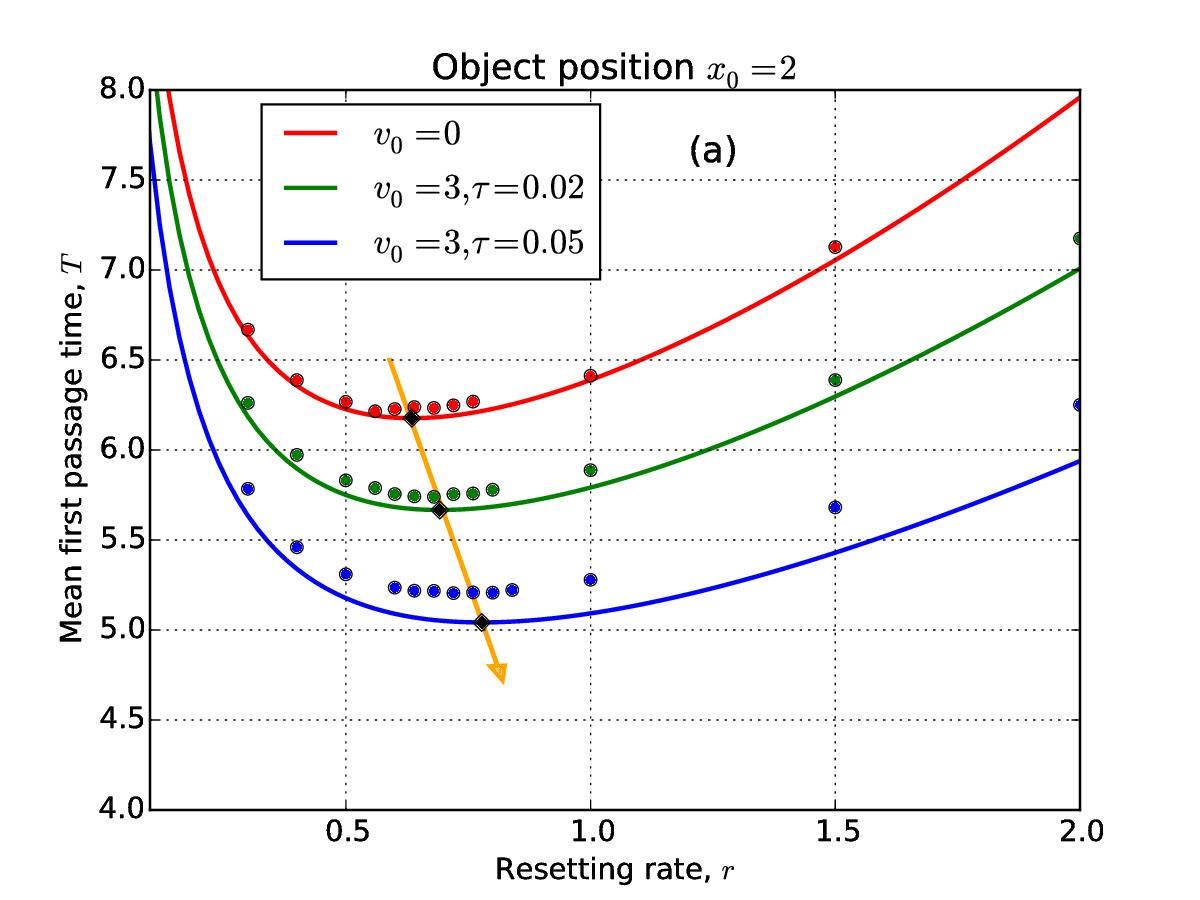} \\
\includegraphics[width=0.53\textwidth]{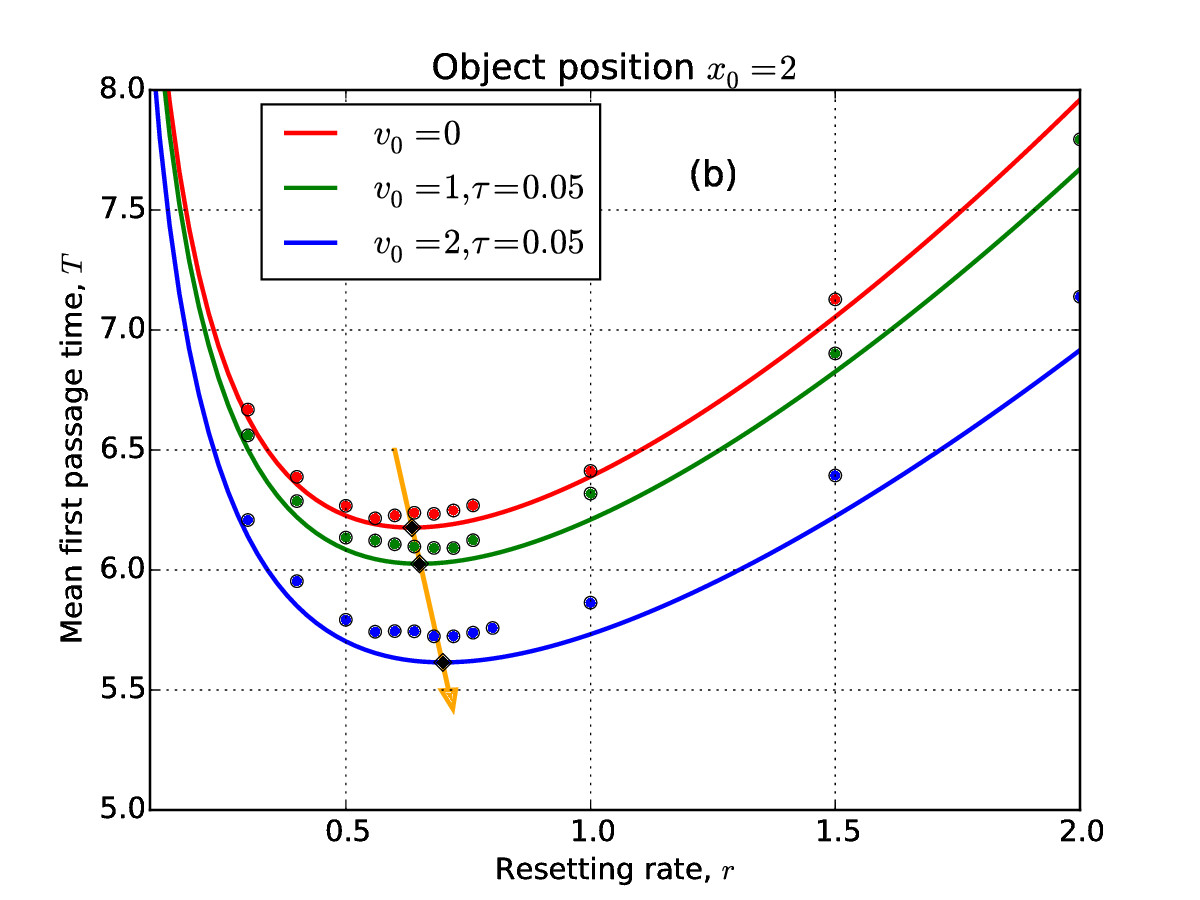}
\end{tabular}
\caption{Mean first passage time in (a) for different persistence times $\tau$ and in (b) for different velocities $v_0$ (see legends). In both figures we also show the results for the Brownian case (red). The lines represent the theoretical predictions of Eq.(\ref{mfpt}) and the filled circles are the simulations results. The black diamonds show the optimal resetting rate (Eq.(\ref{optimal_rate})) and the corresponding MFPT using Eq.(\ref{mfpt}). The arrow indicates the decreasing optimal rate with increasing activity.}
\label{free}
\end{figure}

We first discuss the MFPT for a freely diffusing ABP with resetting. We investigate the effect of activity on the MFPT by considering two cases: (a) fixed value of $v_0 = 3$ and two different values for $\tau=0.02$ and $\tau = 0.05$ in Fig.~\ref{free}(a) and (b) fixed $\tau = 0.05$ and two different values of $v_0 = 1$ and $v_0=2$ in Fig.~\ref{free}(b). In both cases, the target object is located at $x_0 = 2$. In Fig.~\ref{free}, we plot the MFPT as a function of the resetting rate. As expected, the MFPT exhibits a minimum at a particular $r^{\ast}$. As can be seen in Fig.~\ref{free}, the theoretical predictions of Eq.~\eqref{mfpt} are in good agreement with the numerically obtained values.
In order to obtain the optimal resetting rate from simulations, one needs to determine the location of the minimum from the numerical data. However, as can be seen in Fig. \ref{free}, an unambiguous determination of the minimum is difficult in simulations. Nevertheless, on plotting the analytical prediction for the optimal resetting rate (Eq. \eqref{optimal_rate}) together with the numerically obtained data for MFPT, one can clearly see that the Eq. \eqref{optimal_rate} provides an accurate measure for the optimal resetting rate.

\begin{figure}[h!]
\includegraphics[width=0.53\textwidth]{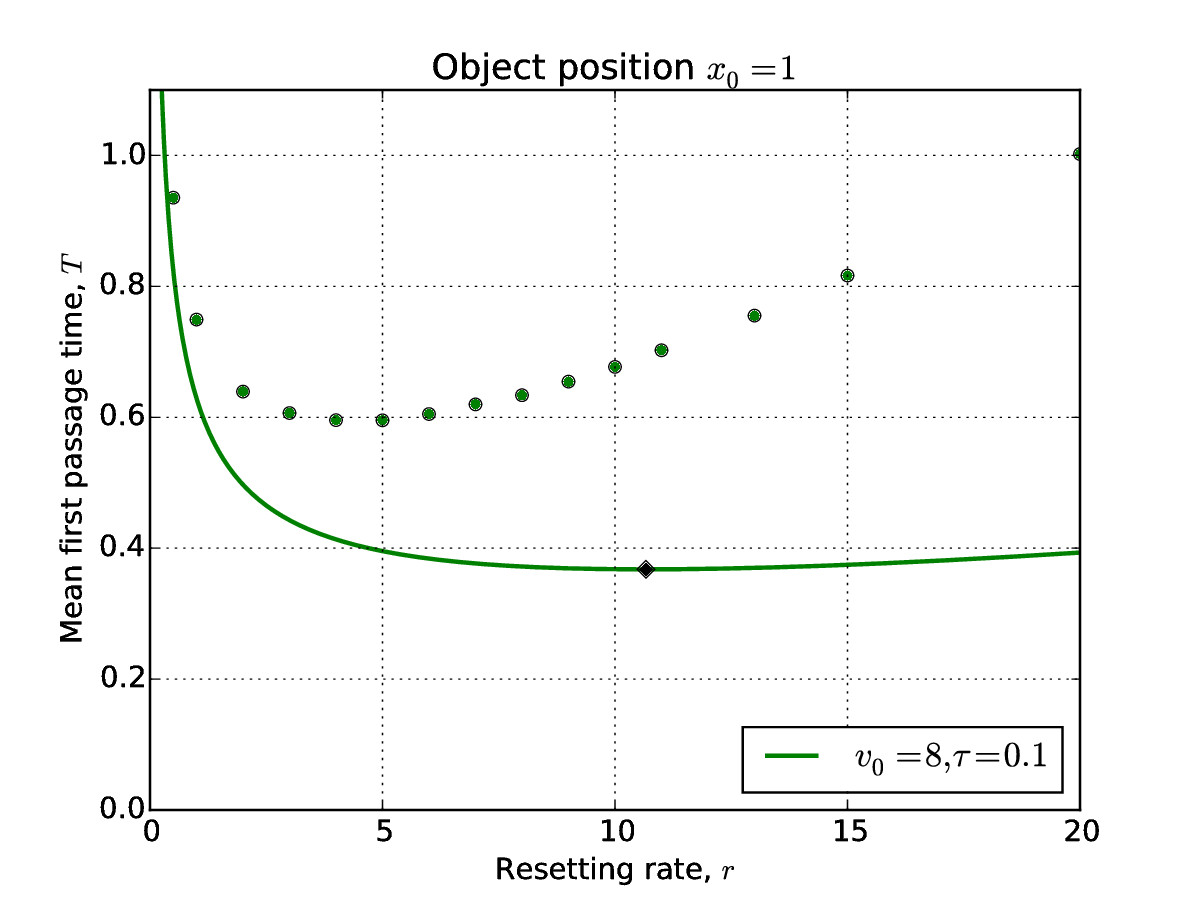}
\caption{MFPT in a case where $x_0\sim v_0\tau$ (see legend). The line represents the theoretical prediction of Eq.(\ref{mfpt}), the black diamond shows the optimal resetting rate from Eq.(\ref{optimal_rate}) with the corresponding MFPT and the symbols are obtained from the numerical simulations. Clearly, the theoretical predictions strongly deviate from the numerical measurements. }\label{bad_params}
\end{figure}

The good agreement between the theoretical predictions of Eqs.~\eqref{mfpt} and~\eqref{optimal_rate} with the numerical results can be qualitatively understood in the following way. The MFPT shown in Fig.~\ref{free} is much larger than $\tau$, the time scale of the persistent motion of an ABP. With $x_0$ chosen to be sufficiently large and resetting rate to be sufficiently small, an active particle diffusing away from the origin will undergo several transitions in its orientation such that the motion of the particle would resemble that of a passively diffusing particle with an effective diffusion constant (Eq.~\eqref{effective_diffusion_coefficient}). However, if $x_0 \sim v_0\tau$, one cannot consider the particle as performing pure diffusion as the discrete nature (in time) of the transitions from one orientation to the other cannot be ignored. In Fig. \ref{bad_params} we show a case with the following choice of parameters: $v_0=8$, $\tau=0.1$ and $x_0 = 1$. These parameters are chosen such that the object can be reached from the origin by the ABP on time scale of the order $\tau$. 
As shown in Fig. \ref{bad_params} the particle reaches the object in a relative short time $t\sim\tau$. On such short time scale the ABP does not undergo enough orientational transitions and therefore does not admit a description in terms of an effective diffusion (passive).

\begin{figure}[h!]
\includegraphics[width=0.53\textwidth]{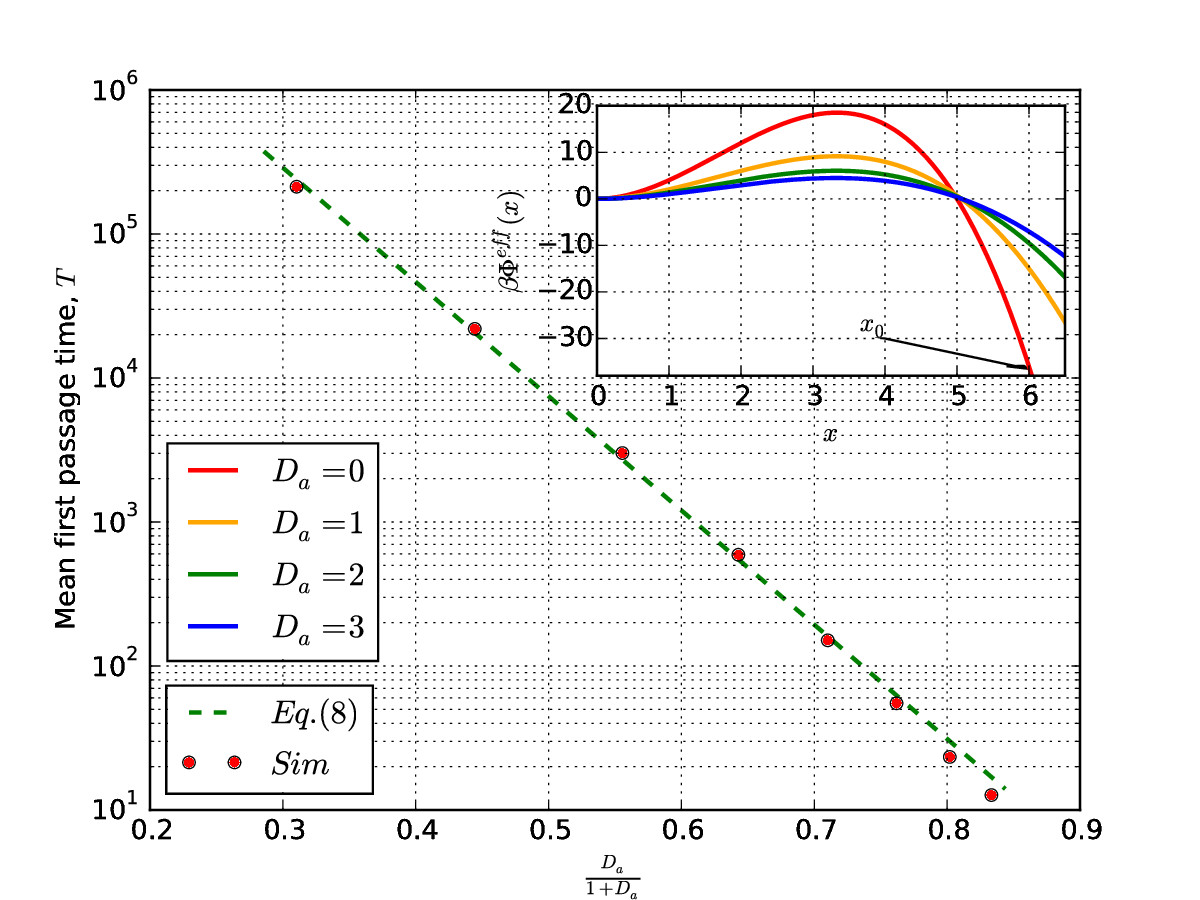}
\caption{Mean first passage time for active particles overtaking an external potential of the form given by equation(\ref{potential}), with the parameters $\alpha=1$ and $\omega_0=10$. The dashed line represents the theoretical prediction using Eq.~(\ref{MFPT_EFF_EQ}) and the symbols are obtained from numerical simulations. There is an excellent agreement between the theoretical predictions and the numerical measurements. The simulations start to deviate from the theoretical results for large $D_a$. In the inset, the bare external potential, Eq.~(\ref{potential}) and analytic effective potentials $\Phi^{eff}(x)$, Eq.~(\ref{Vx}), for the parameter $\tau=0.01$ is shown for different values of $D_a$. The sink is positioned at $x_0=6$, indicated with a black arrow.}\label{log_plot}
\end{figure}

We now discuss the MFPT for an ABP diffusing in an external potential (Eq.(\ref{potential})). We compare the theoretical predictions of Eq.(\ref{MFPT_EFF_EQ}) with the numerical simulations for a specific case where the parameters of the potential are set to be: $\alpha=1$ and $\omega_0=10$. These parameters are chosen to be the same as in Ref. \cite{sharma2017escape}. These numerical results together with the theoretical predictions of Eq. \eqref{MFPT_EFF_EQ} are shown in Fig. \ref{log_plot}. As can be seen in the log-linear plot in Fig. \ref{log_plot}, the MFPT decreases by orders of magnitude with increasing $D_a$, an indication of an exponential dependence on the activity. The exponential dependence of the MFPT on activity is captured in Eq. \eqref{MFPT_EFF_EQ} which is derived in the effective equilibrium approach.  In the effective equilibrium approach, an ABP diffusing in an external potential is mapped to a passive Brownian particle with a modified diffusion constant as well a modified potential. Both the potential and the diffusion constant are determined by the activity parameters. The excellent agreement between the simulations and the theory suggest that a coarse grained description of an ABP diffusing in an external potential is well suited to investigate the MPFT of an ABP. The slight deviation of the numerics from the theoretical predictions for high $D_a$ is due to the fact that for high activities, the modified potential does not satisfy the assumptions (large potential barrier and locally quadratic potential) underlying Eq. \ref{MFPT_EFF_EQ}. This can be seen in the inset of Fig \ref{log_plot} where the effective potential is plotted for different values of $D_a$.

\section{Conclusion}
We investigated the mean first passage time of an Active Brownian Particle in one dimension using numerical simulations. The particle moves with a constant propulsion speed and switches direction at an average rate $\tau$. We studied the influence of a finite resetting rate $r$ on the MFPT to a fixed target of a single free ABP and mapped this result using an effective diffusion process approach. In order to perform analytics, we made the assumption that the motion of an ABP can be described at all time by the ordinary diffusion equation with the diffusion constant given in Eq.~\eqref{effective_diffusion_coefficient}. This is a reasonable assumption only under the self consistency condition that the MFPT is much larger than $\tau$, which is a limitation of this approach. The theoretical predictions are in good agreement with the simulations if the target is located at a distance much larger than the distance travelled by the particle over a period $\tau$, i.e. $x_0 \gg v_0\tau$. As in the case of a passive Brownian particle, we found an optimal resetting rate $r^*$ for an active Brownian particle for which the target is found with the minimum average time. 

In the case of an ABP diffusing in an external potential, we found excellent agreement between the theoretical prediction of the effective equilibrium approach and the computer simulations for ABP. We focussed on a single potential, but similar calculations can be made involving any other trapping potentials. As it was already mentioned in \citep{sharma2017escape}, this excellent agreement suggests that in one dimension, the effective potential can yield accurate quantitative description of the escape process. The validity of the theoretical prediction is limited to low/medium activities.

In this work we focussed on a simple 1-dimensional problem. It will be of interesting to extend this approach to higher dimensions  with the goal of treating more realistic problems.


\section{Acknowledgements}
Alberto Scacchi thanks the Swiss National Science Foundation for financial
support under the grant number 200021-153657/2.\bibliographystyle{apsrev}
\bibliography{references}

%
%

\end{document}